\documentclass[aps,prl,twocolumn,superscriptaddress]{revtex4}
\input epsf
\usepackage{amsmath,amssymb}
\begin{document}

\title{Reversibility of Superconducting Nb Weak Links Driven by the Proximity Effect in a Quantum Interference Device}
\author{Nikhil Kumar}
\affiliation{Department of Physics, Indian Institute of Technology Kanpur, Kanpur 208016, India}
\author{T. Fournier}
\affiliation{Universit\'e Grenoble Alpes, Institut N\'eel, F-38042 Grenoble, France}
\affiliation{CNRS, Institut N\'eel, F-38042 Grenoble, France}
\author{H. Courtois}
\affiliation{Universit\'e Grenoble Alpes, Institut N\'eel, F-38042 Grenoble, France}
\affiliation{CNRS, Institut N\'eel, F-38042 Grenoble, France}
\author{C. B. Winkelmann}
\affiliation{Universit\'e Grenoble Alpes, Institut N\'eel, F-38042 Grenoble, France}
\affiliation{CNRS, Institut N\'eel, F-38042 Grenoble, France}
\author{Anjan K. Gupta}
\affiliation{Department of Physics, Indian Institute of Technology Kanpur, Kanpur 208016, India}
\date{\today}

\begin{abstract}
We demonstrate the role of proximity effect in the thermal hysteresis of superconducting constrictions. From the analysis of successive thermal instabilities in the transport characteristics of micron-size superconducting quantum interference devices with a well-controlled geometry, we obtain a complete picture of the different thermal regimes. These determine whether the junctions are hysteretic or not. Below the superconductor critical temperature, the critical current switches from a classical weak-link behavior to one driven by the proximity effect. The associated small amplitude of the critical current makes it robust with respect to the heat generation by phase-slips, leading to a non-hysteretic behavior.
\end{abstract}


\maketitle

Micron-size superconducting quantum interference devices ($\mu$-SQUID), based on superconducting (SC) weak links (WLs), have been of interest for probing magnetism at small scales \cite{mic-squid-appl,klaus-veauvy-physica,likharev-rmp,dib-apl,denis-nature-nanotech,troeman-nanolett, hao-apl,koshnick-apl}. A major obstacle of a $\mu$-SQUID proper operation is its hysteretic current-voltage characteristic (IVC). During current ramp-up, the WL switches to a dissipative state at the critical current $I_c$ and during current ramp-down, it comes back to a zero-voltage state at the re-trapping current $I_r<I_c$. In conventional tunnel-barrier type Josephson-junctions, the hysteresis arises from large junction capacitance \cite{tinkham-book}. In WLs with negligible capacitance, hysteresis is found at low temperatures below a crossover temperature $T_h < T_c$ \cite{hazra-thesis}, with $T_c$ as the SC critical temperature. Although an effective capacitance can arise from the recovery time of the SC order parameter \cite{song-JAP}, it is now understood that hysteresis in WLs is of thermal origin \cite{hazra-prb,skocpol-jap,tinkham-prb}, similar to that observed in SNS WLs \cite{herve-prl}. A Recent report on high-$T_c$-SC based $\mu$-SQUID shows non-hysteretic IVCs over a wide temperature range \cite{arapia-APL}. Thermal hysteresis in WLs and its effect on IVCs has been modeled by local thermal balance dictating the position of normal metal-superconductor (N-S) interface \cite{skocpol-jap,tinkham-prb,hazra-prb}. In case of poor heat evacuation, phase fluctuations can trigger a thermal run-away giving a resistive hot-spot. This topic is of great practical importance, in particular for SC-magnet wires and cables, helium level sensors, bolometers \cite{bolometer}, $\mu$-SQUIDs and other nano-scale SC structures \cite{shah-nano-wire}. A systematic understanding of various thermal phases which a WL device exhibits can help designing non-hysteretic devices.

In this Letter, we report on the transport characteristics of Nb-film based $\mu$-SQUIDs with a well-controlled geometry and describe a complete picture of different thermal regimes. The IVCs show a critical current and two re-trapping currents that we describe using a thermal instability model in SC leads. The critical current $I_c$ follows the theoretical expectation at low temperature but changes its behavior while crossing the smaller re-trapping current. In this hysteresis-free regime, the WLs superconduct, despite being slightly heated by individual phase slips, thanks to the proximity effect of the adjacent SC.

\begin{figure}
\epsfxsize = 8.1 cm \epsfbox{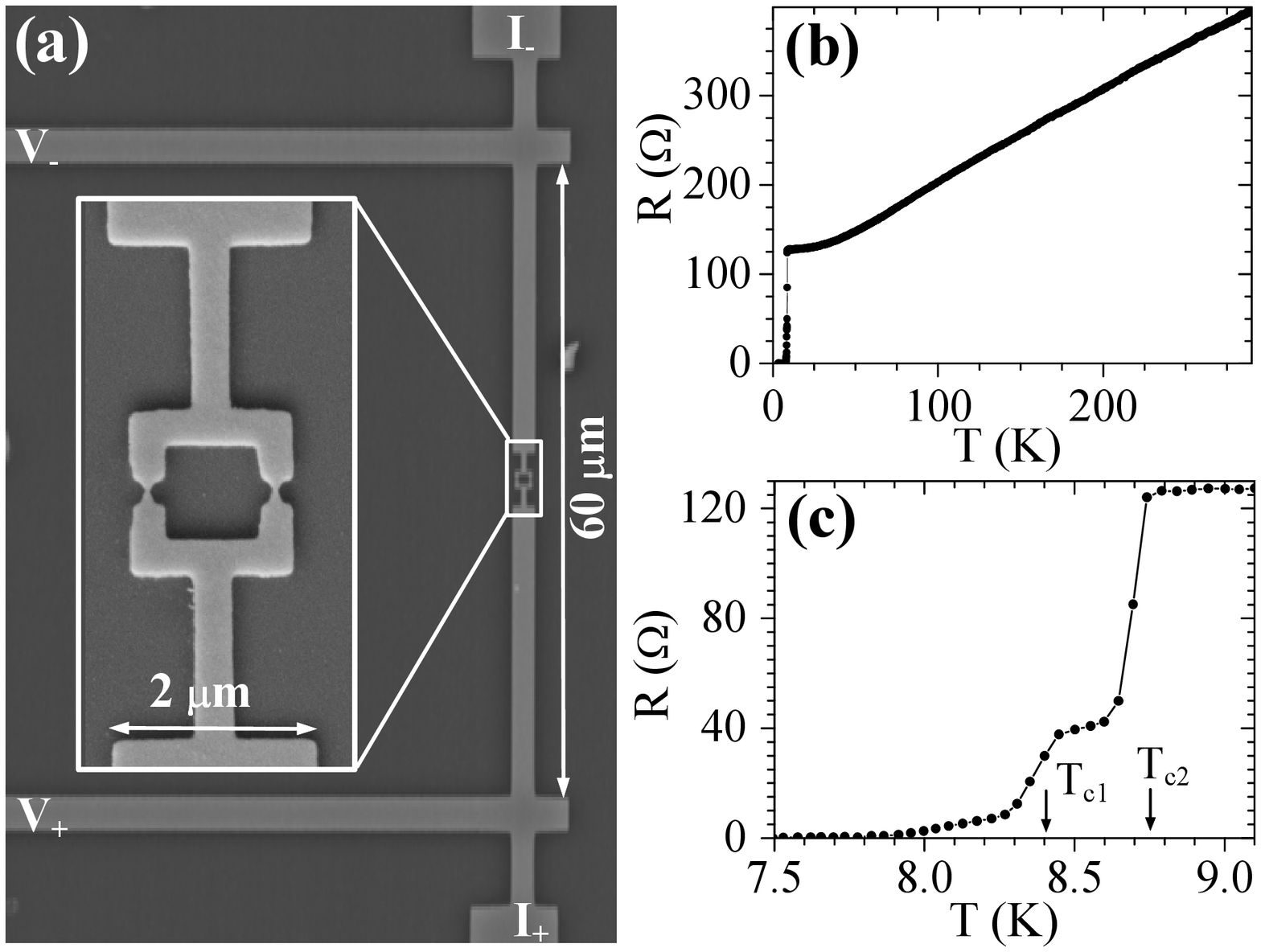}
\caption{(a) SEM image of the $\mu$-SQUID $\mu$S1 with its current and voltage leads. The zoomed-in image shows the SQUID loop (with area 1$\times$1 $\mu$m$^2$) and the narrow leads. (b) Resistance vs temperature (R-T) plot. (c) Low-temperature portion of the R-T plot for $\mu$S1 at 0.01 mA current.}
\label{fig:sem}
\end{figure}

We fabricated \cite{supplementary} $\mu$-SQUIDs from Nb films using common techniques \cite{klaus-veauvy-physica,lam-apl,sophie-prb}. The transport measurements were carried out down to 4.2 K temperature in a homemade cryostat with built-in copper-powder filters \cite{hazra-thesis}. We have studied six devices with similar behavior, but here we report on two devices, $\mu$S1 and $\mu$S2. For all devices, the patterned SQUID-loop area is 1 $\mu$m$^2$ and the width of its arms is 0.3 $\mu$m. The designed WL length is 150 nm while the WL width is 70 and 50 nm in $\mu$S1 and $\mu$S2, respectively. Fig. \ref{fig:sem}(a) shows the SEM image of $\mu$S1. Four different parts of the pattern contribute to the electrical characteristics, namely, 1) the two WLs, each of normal resistance R$_{WL}$, 2) the SQUID loop with normal resistance as R$_L$ including the WLs, 3) the narrow leads of width 0.3 $\mu$m and length 1.7 $\mu$m on either side of the SQUID loop, each with a resistance R$_1$, and 4) the wide leads of width 2 $\mu$m, length 27.5 $\mu$m and normal resistance R$_2$. From the geometry, the total normal-state resistance between the voltage leads is R$_N$ = R$_L$+2R$_1$+2R$_2$ = 40.3R$_{\Box}$+0.5R$_{WL}$. Here, R$_{\Box}$ is the film's square resistance.

Figure \ref{fig:sem}(b) and (c) show the resistance R Vs temperature for $\mu$S1. Multiple SC transitions are observed. The resistance jumps from its residual value of 128 $\Omega$ down to about 40 $\Omega$ at $T_{c2}$ = 8.7 K, jumps further down from 38 to 8 $\Omega$ at $T_{c1}$ = 8.35 K, and finally decreases smoothly to zero. We attribute the transition at $T_{c2}$ to the wide leads and that at $T_{c1}$ to both the narrow leads and the SQUID loop. From IVC in non-hysteretic regime, discussed later [see Fig. \ref{fig:IV}(f)], we deduce R$_{WL}\simeq$ 8 $\Omega$. This analysis is consistent with R$_{\Box}=$ 3.1 $\Omega$, giving a resistivity of 9.5 $\mu\Omega$.cm.

\begin{figure}
\begin{center}
\epsfxsize = 8 cm \epsfbox{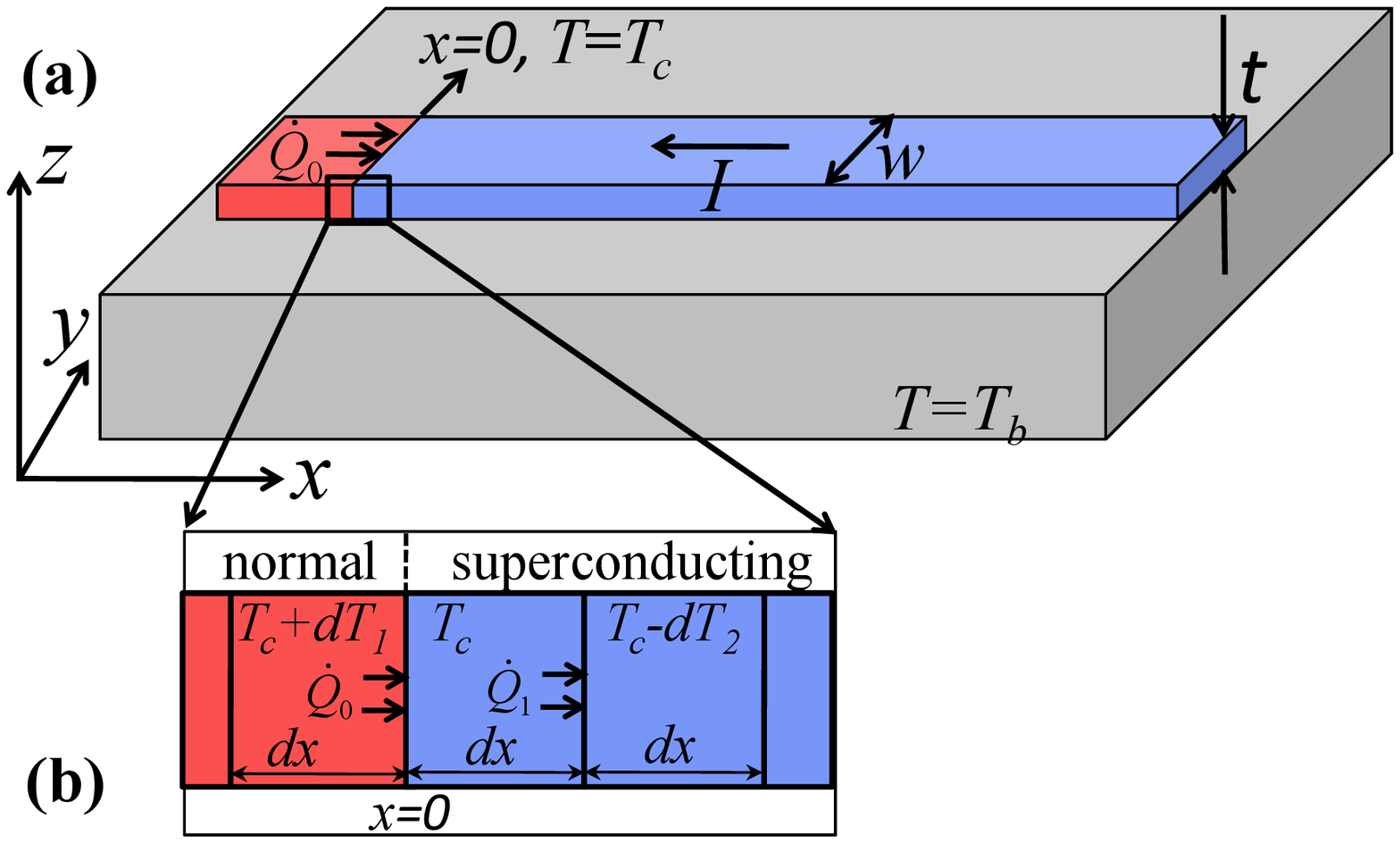}
\caption{(a) Schematic of the semi-infinite (in $+x$ direction) lead of SC material on a substrate at $T_b$ with N-S interface at $x=0$. (b) shows the region near the N-S interface with three differential elements of length $dx$ when the N-S interface stabilizes near the heat source on left.}
\label{fig:lead-config}
\end{center}
\end{figure}

\begin{figure}
\begin{center}
\epsfxsize = 8.5 cm \epsfbox{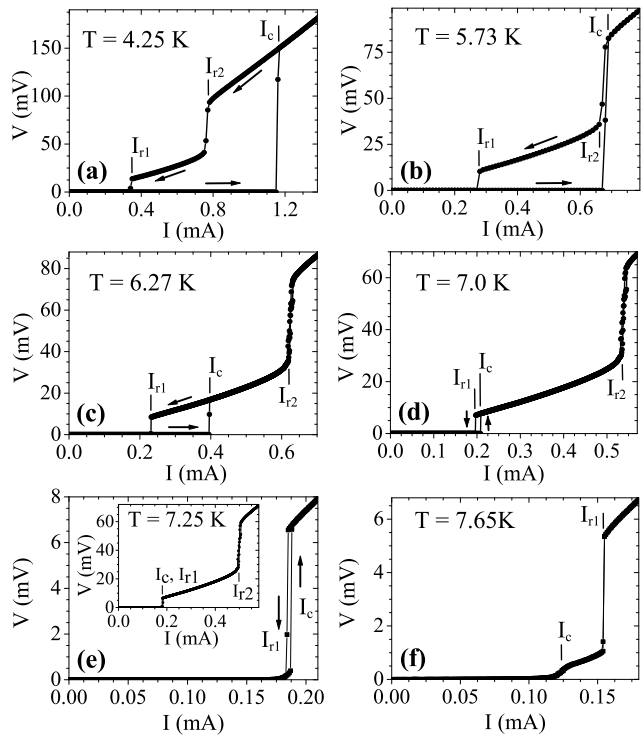}
\caption{(a) - (d) IVCs in hysteretic regime for $\mu$S1 at different temperatures. A large hysteresis is seen at 4.25 K with two re-trapping currents, $I_{r1}$ and $I_{r2}$. $I_c$ crosses $I_{r2}$ near 5.7 K and $I_{r1}$ around 7.25 K as seen in (e). (f) shows the IVC of $\mu$S1 in the non-hysteretic regime above $T_h=$ 7.25 K. The inset of (e) shows a larger bias-current range plot to show the $I_{r2}$ transition.}
\label{fig:IV}
\end{center}
\end{figure}

Next we discuss a one-dimensional model of thermal instability in long current-biased SC leads. This is similar to Broom and Rhoderick \cite{broom-th-stab} model on the dynamics of an N-S interface under the influence of a current. Thus a critical magnitude of current is found at which the N-S interface changes its direction of motion. Here we consider a SC lead with normal state resistivity $\rho_n$, uniform thickness $t$ and width $w$, and carrying an electrical current $I$ as shown in Fig. \ref{fig:lead-config}(a). The heat transfer with the substrate at a bath temperature $T_b$ writes $\alpha (T-T_b)/t$, where $\alpha$ is a characteristic of the interface. The thermal conductivity $\kappa$ is constant and uniform. An N-S interface exists at $x=0$, so at this point $T=T_c$. A heat current flows from $x<0$ due to the resistance of this lead portion plus possibly a device at the end of the lead. With the boundary condition $T=T_b$ at $x\rightarrow\infty$, the heat equation solution for $x>0$ is $T=T_b+(T_c-T_b)\exp(-x/l_{th})$. The thermal healing length $l_{th} (=\sqrt{\kappa t/\alpha})$ is a crossover length-scale such that for $\Delta x \gg l_{th}$ substrate heat-loss dominates and for $\Delta x \ll l_{th}$ conduction dominates. The heat current at the N-S interface ($x=0$) is then $\dot{Q_0}=w\alpha l_{th}(T_c-T_b)$ implying an effective thermal resistance of $(w\alpha l_{th})^{-1}$ as seen from the N-S interface. It is important to realize that the N-S interface will shift to the right (left) if more (less) than $\dot{Q_0}$ heat is incident on the lead at $x=0$.

For analyzing the stability of the N-S interface, we look into the effect of fluctuations on a differential element (from $x=0$ to $x=dx$) at this interface in a quasi-static approximation. If this element turns resistive, see Fig. \ref{fig:lead-config}(b), an additional power $I^2 \rho_n dx/(wt)$ is generated, which is shared equally between the left and right interfaces to the lead and the substrate receives a negligible amount \cite{supplementary}. The heat current across the new N-S interface is $\dot{Q'}_1=\dot{Q}_0-\alpha(T_c-T_b)wdx+I^2 \rho_n dx/(2wt)$. As pointed out before, if this heat is more (less) than $\dot{Q}_0$, the N-S interface will shift to the right (left) implying instability (stability). Thus the maximum current that the lead can carry without causing a thermal instability is given by,
\begin{eqnarray}
I_{max}=w\sqrt{2\alpha(T_c-T_b)/R_{\Box}}.
\label{eq:max-line-curr}
\end{eqnarray}
This expression is consistent with Ref. \cite{skocpol-jap} results in long lead limit and equal thermal conductivities of SC and normal metal, which is valid close to the N-S interface. When $I$ exceeds $I_{max}$, the N-S interface will runaway to a large $x$ location where the lead joins a thermal bath (or a much wider lead). By analyzing the stability of a small resistive element against an incursion to the SC state, one finds as expected the same expression for the re-trapping current. It would be more appropriate to call $I_{max}$ as the `instability current' as it describes both the runaway and re-trapping of the N-S interface. We will use the term `re-trapping' current, as it has been done in most earlier works.

In order to quantify the relevant parameters, we use the Wiedemann-Franz law, i.e. $\kappa=LT/\rho$ with $L=$ 2.44$\times$10$^{-8}$ W.$\Omega$/K$^2$ as the Lorenz number, and using $T=T_c=$ 8.5 K and $\rho=$ 9.5 $\mu\Omega$.cm, we get $\kappa=$ 2.4 W/m.K. Typical values of $\alpha$ used in literature \cite{hazra-prb, skocpol-jap} range from 1 to 10 W/cm$^2$.K. We use $\alpha=$ 5.3 W/cm$^2$.K as found from the temperature dependence of a re-trapping current as discussed later. Thus we find $l_{th}$ = 1.6 $\mu$m for our devices, which is much smaller than the length of the wide leads and comparable to that of the narrow leads.

IVCs of $\mu$S1 in Fig. \ref{fig:IV} shows sharp jumps in voltage at three currents, namely $I_{r1}$, $I_{r2}$ and $I_c$. The jump at $I_c$ occurs during the current ramp up from zero with a distribution in its value. Thus for $\mu$S1 $I_c$ has a width \cite{supplementary} of about 40 $\mu$A with a mean value of 1.3 mA at 4.2 K, in agreement with the expected de-pairing current \cite{depairing-curr}. From the IVC slope, the resistance just above $I_{r1}$ is about 48 $\Omega$. This value is close to the sum R$_L$+2R$_1$ = 40 $\Omega$, which means that the SQUID loop and the narrow leads are heated to above T$_c$ for $I>I_{r1}$. The observed higher value indicates that a portion of the wide leads is also heated to above its $T_c$. The IVC slope above the second re-trapping current $I_{r2}$ is 140 $\Omega$, which is close to the measured residual resistance value, i.e. 128 $\Omega$, indicating a thermal runaway till the voltage leads. The slightly larger value seen here is due to the heating in the central portion to more than 50 K as estimated from a thermal model. At higher temperatures when $I_{r2}$ is much less, indicating reduced heating, the slope above $I_{r2}$ is found to be 128 $\Omega$. In this regime, Fig. \ref{fig:IV}(f) shows that the resistance just above $I_c$ is about 4 $\Omega$ giving R$_{WL}$ = 8 $\Omega$. Only the critical current $I_c$ was found to oscillate with the magnetic flux \cite{supplementary} as expected for a SQUID. The retrapping currents $I_{r1,2}$ do not, implying a different origin than the SC of the WL.

The three currents $I_{r1}$, $I_{r2}$ and $I_c$ evolve differently with temperature. Near 5.7 K, $I_c$ crosses $I_{r2}$ [see Fig. \ref{fig:IV}(b)] and at $T=T_h=$ 7.25K, $I_c$ crosses $I_{r1}$ [see Fig. \ref{fig:IV}(e)], so the hysteresis is absent at higher temperatures [see Fig. \ref{fig:IV}(f)]. In non-hysteretic regime above $T_h$, the IVC near $I_c$ becomes relatively smooth while the voltage jump at $I_{r1}$ remains sharp and evolves over this smooth feature. Also, the hysteresis does not disappear till $I_{r1}$ fully crosses this smooth feature [see Fig.\ref{fig:IV}(e)].

\begin{figure}
\epsfxsize = 8 cm \epsfbox{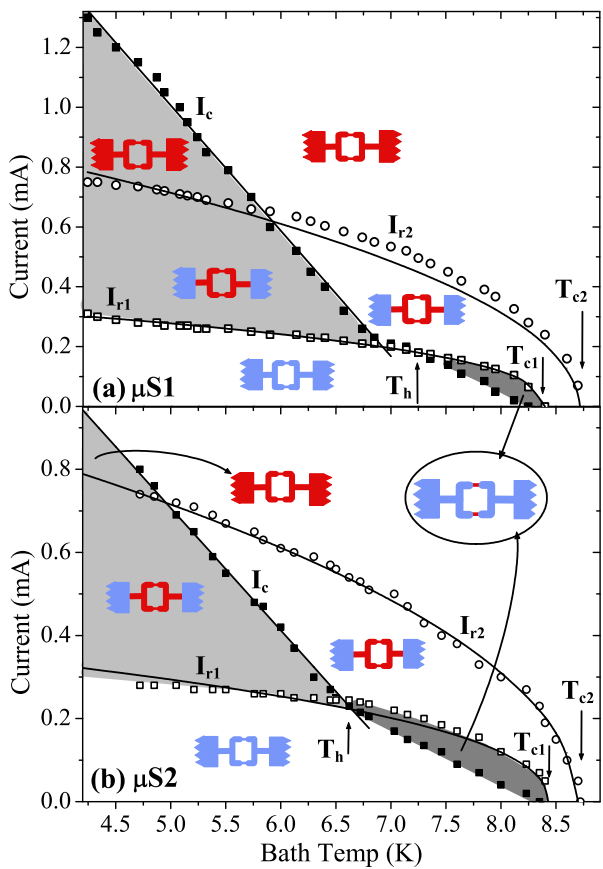}
\caption{Variation of $I_c$, $I_{r1}$ and $I_{r2}$ with $T_b$ for \textbf{(a)} $\mu$S1 and \textbf{(b)} $\mu$S2. The symbols are the data points. The continuous lines are fits given by (in mA and K), \textbf{(a)} $I_c=0.42(7.4-T_b)$ and \textbf{(b)} $I_c=0.29(7.4-T_b)$ while the other two are described by $I_{r1}=0.17(8.4-T_b)^{0.43}$ and $I_{r2}=0.37(8.7-T_b)^{0.5}$ for both the devices. The cartoon pictures of the device shown in different regions depict the state of the device during current ramp-down with blue as SC and red as resistive portions. The light gray-shaded area shows the bistable region where the whole device is in the fully SC state during the current ramp-up from zero. In the dark gray-shaded region, only WLs are resistive.}
\label{fig:IcIrT}
\end{figure}

Figure \ref{fig:IcIrT}(a) summarizes the bath-temperature dependence of $I_c$, $I_{r1}$ and $I_{r2}$ for $\mu$S1. Fig. \ref{fig:IcIrT}(b) shows the same for the device $\mu$S2, with a smaller critical current, and thus a smaller $T_h$. The retrapping currents $I_{r1,2}$ are the same in the two samples confirming that these are independent of the WL structure. With increasing $T_b$, $I_c$ decreases linearly in both devices up to $T_h$, where it shows a marked change in behavior. For both devices, $I_c$ and $I_{r1}$ go to zero at $T_{c1}$, while $I_{r2}$ vanishes at $T_{c2}$. This is consistent with the R-T behavior of Fig. \ref{fig:sem}(c). In both plots, we also indicate the state (resistive or SC) of different portions of the device when the current is ramped down, which constitutes a kind of a phase-diagram, or more appropriately, state-diagram. The light gray-shaded area shows the bistable region where the whole device is SC during the current ramp-up from zero. In the dark gray-shaded region, only the WLs are resistive. No hysteresis is observed in the related temperature range [$T_h$,$T_{c1}$]. This is the most desirable mode for a SQUID, but it occurs in quite a limited temperature window. At a fixed current bias, we do see the expected voltage oscillations with flux in this regime \cite{supplementary}.

Using the long lead approximation for the wide leads, we can fit $I_{r2}$ with Eq. \ref{eq:max-line-curr} which writes here $I_{r2}=w\sqrt{2\alpha(T_{c2}-T_b)/R_{\Box}}$. We obtain a very good fit, see Fig. \ref{fig:IcIrT} with the only free parameter being $\alpha=$ 5.3 W/cm$^2$.K, in good agreement with reported values \cite{hazra-prb,skocpol-jap}. With the same parameters, except $w=$ 0.3 $\mu$m, Eq. \ref{eq:max-line-curr} predicts for the narrow leads a current $I_{r1}$ significantly smaller than observed. This is expected as the presence of wide leads at a short distance makes the heat evacuation more efficient, leading to a higher run-away current.

In a WL with dimensions less than the SC coherence length, we expect, close to its $T_c$, $I_c R_{WL}=\beta(T_c-T_b)$ with $\beta=$ 0.635 mV/K \cite{likharev-rmp}. Our devices are in the Josephson regime \cite{likharev-rmp}, at least close to T$_c$. From the $I_c$ slope in Fig. \ref{fig:IcIrT}(a) for $\mu$S1 at temperatures below $T_h$, we find a $R_{WL}$/2 value of 3 $\Omega$, which agrees with our earlier findings. In this same regime, the extrapolated $T_c$ value of 7.4 K is related to the intrinsic SC of the WLs. Above $T_h$, the $T_b$ dependence of $I_c$ changes slope and $I_c$ goes to zero precisely at $T_{c1}$. Hence we conclude that the WLs are SC above $T_h$ owing to proximity effect from the adjacent SC with a higher $T_c$.

Finally, we elaborate on how the behavior change of $I_c$ coincides with $T_h$. Below $T_h$, $I_c$ exceeds the stability current $I_{r1}$. In this case, even a single phase-slip event induced by thermal fluctuations can cause a thermal runaway \cite{shah-nano-wire}. IVCs thus exhibit a sharp voltage jump at $I_c$ with a distribution in $I_c$ values \cite{fulton-dunkel} because the transition is caused by stochastic fluctuations. Above $T_h$, $I_c<I_{r1}$, so that no thermal runaway can happen at $I_{c}$: the reversible (mono-stable) regime is obtained. The transition to the resistive state (at $I_c$) is smeared with a finite voltage below $I_c$, see Fig. \ref{fig:IV}(e). This is due to phase-slip proliferation as the energy barrier for phase-slip is small for currents close to $I_c$ \cite{fulton-dunkel}. The related dissipation just below $I_c$ also heats some portion of the device above $T_b$. Assuming that the whole SQUID loop is at nearly uniform temperature, which is justified since its size is comparable to $l_{th}$, we estimate that the power generated just below $I_c$ of 72 nW for $T_b$ = 7.25 K brings the SQUID loop to a temperature of about 7.8 K. Because of this and of the fact that the WL region is actually a SC with a lower $T_c$, the $T_b$ dependence of $I_c$ between $T_h$ and $T_{c1}$ cannot be simply described by that of S-N-S WLs \cite{herve-SNS-prb}. Nevertheless, close to $T_c$ we can expect a linear temperature dependence as is the case with both SNS WL and constriction \cite{likharev-rmp}. The heating will reduce $I_c$ value and the exact temperature dependence, close to $T_c$, would be sub-linear.

$I_c$ and $I_{r1}$ are expected to cross at some temperature even if the WL $T_c$ is same as that of the adjacent SC. But then the reversible regime will exist over a narrower temperature range. Thus the smaller $T_c$ of the WL and the proximity SC plays crucial role in widening this hysteresis-free temperature range. By reducing the width of the constriction while keeping other dimensions same one can reduce $I_c$ without affecting $I_{r1}$. This will definitely widen the temperature range of reversible operation. Although at extremely low temperatures, due to divergent Kapitza resistance making $\alpha$ approach zero, the hysteresis is expected to occur even for very small $I_c$. This regime is yet to be investigated.

In conclusion, we present the complete device-state diagram of Nb based $\mu$-SQUIDS. We highlight a non-classical weak link behavior which is understood in the framework of a thermal instability picture. The non-hysteretic high temperature regime of the weak-links is shown to benefit from proximity superconductivity. The present new understanding of the physical mechanisms at the origin of a non-hysteretic behavior is key to further developments in $\mu$-SQUID magneto-sensors for which the suppression of hysteresis represents a key issue.

Samples were fabricated at the platform Nanofab, CNRS Grenoble and measurements were carried out in IIT Kanpur. AKG thanks University Joseph Fourier for a visiting fellowship. NK acknowledges the financial support from CSIR, India. This work has been financed by the French Research National Agency, ANR-NanoQuartet (ANR12BS1000701) and the CSIR of the govt. of India.

\section*{Supplementary Information}
\subsection{Fabrication Details}

We fabricated $\mu$-SQUIDs from Nb films using electron beam lithography. After cleaning the Si substrate with an oxygen plasma, we deposited a 31 nm thick Nb film using e-beam evaporation in a UHV system. We then patterned the structures with electron beam lithography followed by deposition of a 20 nm thick Al film. A lift-off then transferred the pattern to the Al film, which acts as a mask during the reactive ion etching of Nb using SF$_6$ plasma. Finally, the Al film was removed chemically.

\subsection{SQUID oscillations with magnetic flux}
\begin{figure}[h]
\epsfxsize = 3.2 in \epsfbox{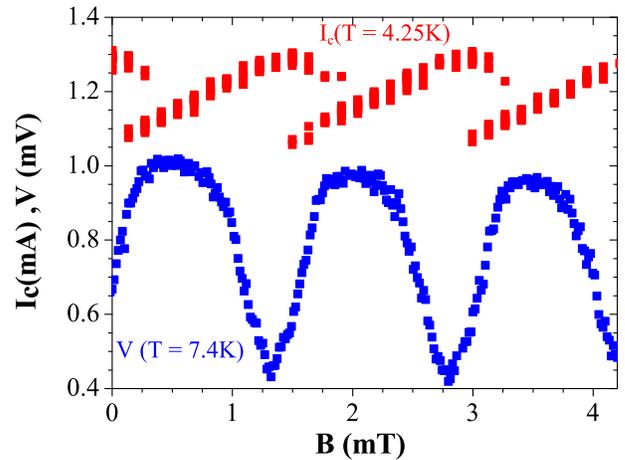}
\caption{$I_c$ oscillations in hysteretic regime for $\mu$S1 at 4.25K (Red curve) and the voltage oscillations (at 0.17mA current) in non-hysteretic regime at 7.4K (Blue curve).}
\label{fig:Ic-V-vs-B}
\end{figure}
Figure \ref{fig:Ic-V-vs-B} below shows the oscillations in $I_c$ for $\mu$S1 at 4.25 K (below $T_h$) and in voltage at 7.4 K (
above $T_h$) with external magnetic flux. The voltage oscillations are acquired at a bias current of 0.17 mA, which is close to the critical current at this 7.4 K. The SQUID oscillations with magnetic field are seen only in $I_c$ and not in $I_{r1}$ and $I_{r2}$. The temperature dependent $I_c$ values have been extracted from these $I_c$-Vs-B plots at all temperatures by selecting maximum $I_c$ at each temperature. In the non-hysteretic regime $I_c$ was found from the maximum slope of the IVC. This is found to coincides with the current at which the voltage modulation in V-Vs-B peaks. In both cases, the magnetic field periodicity is found to be 1.5 mT, which defines an effective SQUID loop area as $A_{eff} = \frac{\phi_0}{\Delta B}= 1.3 \mu m^2$, which is larger than the actual patterned (internal) area of 1 $\mu m^2$.

\subsection{Heat sharing during resistive fluctuation}
In order to elaborate on the sharing of the extra resistive heat, when the differential element becomes normal, by the three interfaces, we also consider two neighboring differential elements of the same length $dx$ as shown in Fig. 2(b) of the main paper. The one on the left (i.e. from $x=-dx$ to $x=0$) is at temperature $T_c+dT_1$ and the one on the right (i.e. from $x=dx$ to $x=2dx$) is at temperature $T_c-dT_2$. The left one gives heat $\dot{Q}_0=\kappa w t \frac{dT_1}{dx}$ to the middle one, which gives heat $\dot{Q}_1=\kappa w t \frac{dT_2}{dx}$ to the element on right and thus we get,
\begin{eqnarray}
\kappa w t \frac{dT_1}{dx}=\alpha(T_c-T_b) w dx+\kappa w t \frac{dT_2}{dx}
\label{eq:heat-bal-eqbm}
\end{eqnarray}
When the middle element becomes resistive due to fluctuations its temperature increases to $T_c+dT$. In this case the above equation gets modified to
\begin{eqnarray}
\kappa w t \frac{dT_1-dT}{dx}&=&\alpha(T_c+dT-T_b) w dx \nonumber\\ &+&\kappa w t \frac{dT_2+dT}{dx}-I^2\frac{\rho_n}{wt}dx
\label{eq:heat-bal-fluct}
\end{eqnarray}
Subtracting eq. \ref{eq:heat-bal-eqbm} from eq. \ref{eq:heat-bal-fluct} we get $2\kappa w t\frac{dT}{dx}=I^2\frac{\rho_n dx}{wt}-\alpha dT w dx$. Neglecting the higher order second term on the right, we get $\kappa w t\frac{dT}{dx}=I^2\frac{\rho_n dx}{2wt}$. Thus the heat current incident from the left interface, i.e. $\dot{Q'}_0=\kappa w t \frac{dT_1-dT}{dx}=\dot{Q}_0-I^2\frac{\rho_n dx}{2wt}$ and the heat current incident at the right interface, i.e. $\dot{Q'}_1=\kappa w t \frac{dT_2+dT}{dx}=\dot{Q}_1+I^2\frac{\rho_n dx}{2wt}$. Thus the extra heat generated is equally shared across the two interfaces.

\begin{thebibliography}{99}
\bibitem{depairing-curr} We get the depairing current \cite{tinkham-book}, $I_d = 2A(2/3)^{3/2} [Hc(0)/\lambda]$ as 1.8 mA at 0 K by using penetration depth ($\lambda$) as 200nm, thermodynamic critical field [$\mu_0H_c$(0)] as 0.2 T and cross-section area ($A$) for each WL as 31$\times$70 nm$^2$.
\end{thebibliography}
\end{document}